\documentclass[a4paper]{jpconf}
\usepackage{graphicx}

\usepackage{cases}
\usepackage{amsthm}
\usepackage{amsfonts}
\usepackage{amssymb}
\usepackage{multicol}
\usepackage{pdfpages}

\begin{document}

\def\br{\begin{eqnarray}}
\def\er{\end{eqnarray}}

\newtheorem{conjectura}{Conjecture}
\newtheorem{definition}{Definition}
\newtheorem{lema}{Lema}
\newtheorem{theorem}{Theorem}
\newtheorem{proposition}{Proposition}

\title{Rational solutions from Pad\'e  approximants for the generalized 
Hunter-Saxton equation}

\author{ H Aratyn $^{1}$, J F Gomes $^{2}$, D V Ruy $^{2}$ and A H Zimerman $^{2}$}

\address{ $^1$
Department of Physics, University of Illinois at Chicago, 845 W. Taylor St., Chicago, IL 60607-7059}
\address{$^2$Instituto de F\'isica Te\'orica-UNESP, Rua Dr Bento Teobaldo Ferraz 271, Bloco II, 01140-070, S\~ao Paulo, Brazil}

\ead{jfg@ift.unesp.br}

\begin{abstract}
Exact rational solutions of the generalized Hunter-Saxton 
equation are obtained using  Pad\'e approximant approach for the traveling-wave and self-similarity reduction.
A larger class of algebraic solutions are also obtained by extending a range of parameters within the solutions obtained by this approach.

\end{abstract}

\section{Introduction}
\label{section:intro}

The generalized Hunter-Saxton (HS) equation was originally proposed in  \cite{Pavlov}  as
\begin{equation}
u_{xt}=uu_{xx}+k u_x^2. \label{1}
\end{equation}

This model reproduces the original HS equation when $k=1/2$, which is a model for wave propagation in a director field of nematic liquid crystals. Besides, equation (\ref{1})  appears in different applications,  such as the study of the Einstein-Weyl space for $k=-1/3$  \cite{Tod}. 
Solutions of (\ref{1}) in a non-closed form were obtained in
ref. \cite{Pavlov} by a hodographic transformation, however the task of expresing these solutions in a closed form of the original variables is complicated and indirect.

It is known that the Pad\'e approximant usually gives a good representation of the solution near a regular point (see \cite{Baker} for more details) and it has widely been used to construct approximative solutions of differential equations with already determined initial conditions \cite{Boyd,Mousa}. Moreover, an exact solution was found for the Boussinesq equation in \cite{Mousa} for a predetermined initial condition. In this paper we employ the inverse path for study the generalized HS equation. We start with a rational expression given by the Pad\'e approximant and seek  for a set  parameters or  initial conditions that allows the ansatz to become an exact solution. Although there is no guarantee that a given ODE has rational or
polynomial solutions, this approach provides a useful 
tool for searching for them.

The Pad\'e approximant
consists of a ratio of two polynomials that can be expressed as
\[
[L/M](x) \equiv {P_L(x)\over Q_M(x) }={\sum_{j=0}^L p_j x^j\over 1+\sum_{j=1}^M q_j x^j}
\]
where the polynomials $P_L(x)$ and $Q_M(x)$ are defined such that
$[L/M](x)$ agrees with the Taylor expansion a function $f=f(x)$  up to degree $L+M$, i.e.
\[ 
f(x)={P_L(x)\over Q_M(x) }+{\cal O}(x^{L+M+1}). 
\]

Thus, in order to determine the Pad\'e approximant of a Taylor expansion at $x=0$ given by $f=\sum_{j=0}a_j x^j$, we 
need to solve a system of $L+M$ equations, namely, 
\[ \sum_{m+n=j}a_m q_n-p_j=0, \hspace{1
cm} j=0,1,...,L+M . \]

Consider now an arbitrary ordinary differential equation (ODE):
\begin{equation}\label{ODE}
E(z,v,v_z,...;{\cal S}_0)=0 ,   \hspace{1 cm}  
{\cal S}_0 \equiv 
\textnormal{Cartesian product of the sets of parameters},
\end{equation}
and let $v=v(z;{\cal S}_0\times{\cal S}_1)$ be the 
general solution of this equation, where ${\cal S}_1$ is the 
Cartesian product of the sets of initial conditions. Let 
us assume the origin has a regular local representation, i.e. 
$v=\sum_{j=0}^\infty v_j z^j$. Substituting in the differential equation
(\ref{ODE}), we have
\[
E(v;{\cal S}_0)=\sum_{j=0}^\infty E_j(v_0,...,v_{j+n}) z^j=0,
\]
where $n$ is the order of the equation. The above expression determines the coefficients $v_j$ such that
\[
v=\sum_{j=0}^\infty v_j({\cal S}_0\times{\cal
S}_1) z^j \, .
\]

Calculating the Pad\'e approximant of this Taylor expansion, we have
\[
v={P_L(z;{\cal S}_0\times{\cal
S}_1)\over Q_M(z;{\cal S}_0\times{\cal S}_1) }+{\cal O}(z^{L+M+1}) \,.
\]

Let us assume now  that there are subsets $\hat{{\cal S}}_0\subset
{\cal S}_0$ and $\hat{{\cal S}}_1\subset {\cal S}_1$, such that the
Pad\'e approximant of $\hat{{\cal S}}=\hat{{\cal S}}_0\times\hat{{\cal S}}_1$ leads to an exact solution, i.e.
\begin{equation}\label{ansatz}
v(z;\hat{{\cal S}})={P_L(z;\hat{{\cal S}})\over Q_M(z;\hat{{\cal S}}) }. \label{3}
\end{equation}

This ansatz allows us to seek solutions in the rational 
form without a need to consider further properties of 
the differential equation. Substituting (\ref{ansatz}) 
in (\ref{ODE}), we have
\begin{equation}\label{expansion E}
E(z,v,v_z,...;\hat{{\cal S}}_0)={\sum_{j=0}^{\Lambda_{L,M}}\hat{E}_j(\hat{{\cal S}}) z^j \over D(\hat{{\cal S}}) }=0,
\end{equation}
where the degree $\Lambda_{L,M}$ depends of $L$, $M$ and the particular form of the differential equation. In order to determine all elements $s_j\in\hat{{\cal S}}$ for which the ansatz (\ref{ansatz}) is true, we need to solve a system of algebraic equations,
\begin{eqnarray}
 \hat{E}_j (\hat{{\cal S}}) &=& 0   ,  \hspace{1 cm}  j=0,1,...,L+M-n  \label{cond 1}  \\
 \hat{E}_j (\hat{{\cal S}}) &=& 0   ,  \hspace{1 cm}  j=L+M-n+1,...,\Lambda_{L,M}   \label{cond 2}   \\
 D(\hat{{\cal S}}) &\neq& 0   \label{cond 3}  
\end{eqnarray}

Observe that the system (\ref{cond 1}) is identically satisfied by the
definition of the Pad\'e approximant. Thus, the exact  solution (\ref{3}) is determined by the choice of parameters $\hat {{\cal S}}_0$ and  initial conditions $\hat {{\cal S}}_1$.

In the next section, we apply this
approach to the traveling-wave and self-similarity reductions of the
generalized Hunter-Saxton equation. Moreover, for these specific
reductions, we are able to extend the rational solutions obtained by
the Pad\'e approximant method to a more general set of parameters.

\section{Generalized Hunter-Saxton equation}\label{HS}

\subsection{Traveling-wave reduction of the generalized HS equation}

Applying the traveling-wave reduction
\begin{equation} \label{twred}
z= x+\mu t   ,   \hspace{1 cm}  u(x,t)=u(z),\qquad \mu = \rm{constant}
\end{equation}
to the generalized Hunter-Saxton equation  (\ref{1}) yields:
\begin{equation}\label{trav}
-\mu u_{zz}+uu_{zz}+k u_z^2=0 \, .
\end{equation}
The Taylor expansion of a solution to the above equation near the origin ($z=0$) can be
written as
\[
u=u_0+u_1z+{ku_1^2\over2(\mu-u_0)}z^2+{k(2k+1)u_1^3\over
6(\mu-u_0)^2}z^3+{k(6k^2+7k+2)u_1^4\over 24(\mu-u_0)^3}z^4+...\,     \hspace{0.7 cm}  u_0\neq \mu ,
\]
where $u_0$ and $u_1$ are arbitrary constants. In terms of the
formalism from the previous section we have ${\cal S}_0=(k,\mu)$ and ${\cal S}_1=(u_0,u_1)$. 

We now apply the Pad\'e approximant approach to the two 
simplest cases $[1/1]$ and $[2/2]$. Let us begin with the ansatz
\begin{equation}\label{Pade trav 1-1}
u\equiv u(z;(k,\mu,u_0,u_1))={P_{1}(z;\hat{{\cal S}})\over Q_{1}(z;\hat{{\cal S}})}
\end{equation}
where
\begin{eqnarray*}
P_{1}(z;\hat{{\cal S}}) &=& u_0+{(2\mu-(k+2)u_0)u_1\over 2(\mu-u_0)}z   \\
Q_{1}(z;\hat{{\cal S}}) &=&  1-{ku_1\over2(\mu-u_0)}z
\end{eqnarray*}

For this ansatz,  the system (\ref{cond 2}) is given by
\begin{eqnarray}
 \hat{E}_1 (\hat{{\cal S}}) &=& 8k(k+2)u_1^3(\mu-u_0)^3=0  ,  \label{cond HS trav 1/1}
\end{eqnarray}
with $\Lambda_{1,1}=1$.
The condition (\ref{cond 3})  reads here as 
\begin{equation}\label{cond HS trav 1/1 2}
D(\hat{{\cal S}}) = k^4u_1^4z^4-8k^3u_1^3(\mu-u_0)z^3+
24k^2u_1^2(\mu-u_0)^2z^2-32ku_1(\mu-u_0)^3z+16(\mu-u_0)^4     \neq 0 \, .
\end{equation}
We see that $k=-2$ and $k=0$  simultaneously solve conditions 
(\ref{cond HS trav 1/1}) and (\ref{cond HS trav 1/1 2}). Thus, by 
substituting these
particular values of parameters on (\ref{Pade trav 1-1}), we obtain 
exact solutions:
\begin{eqnarray} 
k=-2 &;& u={(\mu-u_0)u_0+z\mu u_1 \over \mu-u_0+u_1 z} \label{1stansatz 1}  \\
k=0  &;& u=u_0+u_1 z \, . \label{1stansatz 2}
\end{eqnarray}
Let us now take as an ansatz the Pad\'e approximant of type [2/2] :
\begin{equation}\label{Pade trav 2-2}
u\equiv u(z;(k,\mu,u_0,u_1))={P_{2}(z;\hat{{\cal S}})\over Q_{2}(z;
\hat{{\cal S}})}\,,
\end{equation}
where
\begin{eqnarray*}
P_{2}(z;\hat{{\cal S}}) &=& u_0+{(2\mu-(2k+3)u_0)u_1\over 2(\mu-u_0)}z+{(-6(k+1)\mu-(2k^2+7k+6)u_0)u_1^2  \over 12(\mu-u_0)^2}z^2   \\
Q_{2}(z;\hat{{\cal S}}) &=&
1-{(2k+1)u_1\over2(\mu-u_0)}z+{k(2k+1)u_1^2\over 12(\mu-u_0)^2}z^2 \, .
\end{eqnarray*}
It follows that for this ansatz the system (\ref{cond 2}) is given by
\begin{eqnarray}
 \hat{E}_3 (\hat{{\cal S}}) &=& 576k(k+2)(2k+1)(2k+3)u_1^5(\mu-u_0)^5=0  \label{HS trav 2/2 1}\\
 \hat{E}_4 (\hat{{\cal S}}) &=& -432k(k+1)(k+2)(2k+1)(2k+3)u_1^{6}(\mu-u_0)^4=0  \label{HS trav 2/2 2}  \\
 \hat{E}_5 (\hat{{\cal S}}) &=& 24k(k+2)(2k+1)^2(2k+3)^2 u_1^{7} (\mu-u_0)^3=0    \label{HS trav 2/2 3}
\end{eqnarray}
supplemented by condition (\ref{cond 3}), i. e.
\begin{eqnarray*}
D(\hat{{\cal S}}) &=&
k^4(2k+1)^4u_1^8z^8-24k^3(2k+1)^4u_1^7(\mu-u_0)z^7\\
&+&24k^2(2k+1)^3(20k+9)u_1^6(\mu-u_0)^2z^6-  
864k(2k+1)^3(3k+1)u_1^5(\mu-u_0)z^5\\
&+&432(2k+1)^2(38k^2+24k+3)u_1^4(\mu-u_0)^4z^4 \\
&-& 10368(2k+1)^2(3k+1)u_1^3(\mu-u_0)z^3+3456(2k+1)(20k+9)u_1^2(\mu-u_0)z^2  \\
&-&  41472(2k+1)u_1(\mu-u_0)^7z+20736(\mu-u_0)^8      \neq 0
\end{eqnarray*}
Substituting the solution of system 
(\ref{HS trav 2/2 1}-\ref{HS trav 2/2 3}) into 
the ansatz (\ref{Pade trav 2-2}) yields exact solutions:
\begin{eqnarray*}
k=-2 &;& u={(\mu-u_0)u_0+z\mu u_1 \over \mu-u_0+u_1 z}  \\
&& \\
k=-3/2 &;& u={4(u_0-\mu)u_0+4(\mu-u_0)\mu u_1 z+\mu u_1^2z^2 \over (2(\mu-u_0)+u_1z)^2} \\
&& \\
k=-1/2 &;& u=u_0+u_1 z -{u_1^2z^2 \over 4(\mu-u_0)} \\
&& \\
k=0  &;& u=u_0+u_1 z 
\end{eqnarray*}
Notice that with the above choice of parameters the order of the Pad\'e approximants  is reduced accordingly.

Observe that solutions (\ref{1stansatz 1}) and (\ref{1stansatz 2}) obtained from the first ansatz 
are included in the above set of solutions. This is not a general case as it
will be evident in the next section. These ansatzes can be worked out
by an algebraic manipulation software such as Mathematica or Maple so
we omit details for the system (\ref{cond 1}) and (\ref{cond 2})
in the remaining part of the text. Following similar  procedure as above for 
the ansatz:
\begin{equation}\label{Pade trav 3-3}
u\equiv u(z;(k,\mu,u_0,u_1))={P_{3}(z;\hat{{\cal S}})\over Q_{3}(z;\hat{{\cal S}})},
\end{equation}
we obtain the solutions

\begin{eqnarray*}
 k=-2  &;&   u=\mu+{-\mu^2+2\mu u_0-u_0^2 \over \mu+u_1z-u_0} \\
&&    \\
 k=-3/2 &; &   u=\mu-{4(\mu^3-3\mu^2u_0+3\mu u_0^2-u_0^3) \over (2\mu+u_1 z-2u_0)^2} \\
&&    \\
 k=-4/3  &;&   u=\mu-{27(\mu^4-4\mu^3 u_0+6\mu^2u_0^2-4\mu u_0^3+u_0^4) \over (3\mu+u_1 z-3u_0)^3} \\
&&    \\
k=-{2\over3} &;&  u={u_1^3 z^3\over 27(\mu-u_0)^2}-{u_1^2 z^2\over 3(\mu-u_0)}+u_0+u_1 z   \\
&&     \\
k=-{1\over2} &;&  u=-{u_1^2 z^2\over 4(\mu-u_0)}+u_0+u_1 z   \\
&&    \\
k=0  &;& u=u_0+u_1 z
\end{eqnarray*}

We can see that all non-trivial solutions of the above system are
expressed in terms of  $\hat{{\cal S}_1}={\cal S}_1$ and $\hat{{\cal S}_0}$
being independent of the initial conditions, therefore, they are general
solutions.


A careful study of these solutions suggests
a general expression that compactly  expresses the whole 
set $\hat{{\cal S}}$. 
Indeed direct verification confirms that all solutions  are reproduced by
\begin{equation}\label{solution HS traveling}
u=\mu+{(u_0-\mu)\over (1+{(-1)^r z u_1\over r(-\mu+u_0)})^r},    \hspace{1 cm}   r=-{1\over (k+1)}. \label{19}
\end{equation}
Although this solution was obtained originally for integer values of $r$
only it can be checked that it holds for any $r$ different from $0$. 
 The closed general solution for $k=1/2$ was obtained
recently in \cite{Al-Ali} and agrees with (\ref{19}).


\subsection{Self-similarity reduction of the
generalized HS equation}
In this section, we explore the self-similarity reduction of the
generalized Hunter-Saxton equation. Observe that inserting 
the self-similarity reduction expression
\begin{equation} \label{ssred}
    u(x,t)=t^{-(1+\xi)}v(z),    \hspace{1 cm}    z=xt^\xi, \qquad
    \xi = \rm{constant}
\end{equation}
into (\ref{1}) yields 
\begin{equation}\label{self-sim}
v_z-\xi zv_{zz}+vv_{zz}+kv_z^2=0
\end{equation}
The local representation of the solution $v$ around the origin is
\begin{equation}
v=v_0+v_1z-{v_1(1+kv_1)\over 2v_0}z^2+{v_1(1+kv_1)(1-\xi+(1+2k)v_1)\over 6v_0^2}z^3+...,\label{x}
\end{equation}
where $v_0$ and $v_1$ are arbitrary constants, ${\cal S}_0=(k,\xi)$ and ${\cal S}_1=(v_0,v_1)$. Notice that  for $v_1 = -1/k$ the above series truncates leading to the solution
\[
v=v_0-{{z}\over {k}} .
\]

We can apply  the ansatz for the Pad\'e
approximants provided that $v_1 \neq -{{1}\over {k}}$.  Therefore, considering the ansatz $v\equiv v(z;(k,\xi,v_0,v_1))={P_{1}(z;\hat{{\cal S}})\over Q_{1}(z;\hat{{\cal S}})}$, we have the set $\hat{{\cal S}}$ and the corresponding solution given by
\begin{eqnarray*}
(k,\xi,v_0,v_1)=(-2,-1/2,v_0,v_1) &;&v={v_0(2v_0+z) \over 2v_0+(1-2v_1)z} 
\end{eqnarray*}

Applying the same procedure to the ansatz $v\equiv v(z;(k,\xi,v_0,v_1))={P_{2}(z;\hat{{\cal S}})\over Q_{2}(z;\hat{{\cal S}})}$, it give us
\begin{eqnarray*}
(k,\xi,v_0,v_1)=(-2,1,v_0,v_1) &;& v={2v_0^2-v_1z^2 \over 2v_0-2v_1z} \\
&&\\
(k,\xi,v_0,v_1)=(-3/2,-1/3,v_0,v_1)&;&v={2v_0[18u_0^2+12v_0z+(2-3v_1)z^2] \over [6v_0+(2-3v_1)z]^2}   \\
&& \\
(k,\xi,v_0,v_1)=(-1/2,1,v_0,v_1)&;&v=v_0+v_1 z+{v_1(v_1-2)\over 4v_0}z^2 
\end{eqnarray*}

In order to obtain more solutions which  helps to look for a compact expression of all rational solutions, we also consider the ansatz $v\equiv v(z;(k,\xi,v_0,v_1))={P_{3}(z;\hat{{\cal S}})\over Q_{3}(z;\hat{{\cal S}})}$. This yields the following solutions
\begin{eqnarray*}
(k,\xi,v_0,v_1)=(-3/2,1,v_0,v_1)&;&v={2(6v_0^3-3v_0v_1z^2+v_1^2z^3) \over 3(2v_0-zv_1)^2}  \\
&& \\
(k,\xi,v_0,v_1)=(-4/3,-1/4,v_0,v_1)&;&v=-{3v_0[576u_0^3+432v_0^2z-36v_0(4v_1-3)z^2+(3-4v_1)^2z^3] \over [(4v_1-3)z-12v_0]^3}  \\
&& \\
(k,\xi,v_0,v_1)=(-2/3,1/2,v_0,v_1)&;&v=v_0+v_1 z+{v_1(2v_1-3)\over 6v_0}z^2+{v_1(2v_1-3)^2\over 108v_0^2}z^3  \\
&& \\
(k,\xi,v_0,v_1)=(-2/3,1,v_0,v_1)&;&v=v_0+v_1 z+{v_1(2v_1-3)\over 6v_0}z^2+{v_1^2(2v_1-3)\over 54v_0^2}z^3.
\end{eqnarray*}


We see that the parameters of the above general solutions can be organized as
\[
k=-{(r+1)\over r}   \hspace{1 cm}  \rm{with}   \hspace{1 cm}  \xi=1    \hspace{1 cm} or    \hspace{1 cm}   \xi=-{1\over(r+1)}
\]

In order to  express the
solutions  in a 
compact form, let us define the constants $u_0,z_0$ as
\[
v_0={z_0 v_1\over r},    \hspace{1 cm}   v_1={r\over (r+1)}+{(-1)^r r u_0\over z_0^{r+1}}
\]
when $\xi=1$ and
\[
v_0=\biggl(-{1\over (r+1)}+{v_1\over r}\biggr)z_0,    \hspace{1 cm}   v_1={(-1)^r r u_0\over z_0^{r+1}}
\]
when $\xi=-1/(r+1)$.

The solutions found by the Pad\'e approximant approach for
the self-similarity reduction can be summarized as
\begin{eqnarray}
v={u_0\over (z-z_0)^r}+{z_0+rz\over r+1} ,   \hspace{1 cm} & \textnormal{when} &   \hspace{1 cm}  \xi=1 ,  \label{general self-similar}   \\
v={u_0\over (z-z_0)^r}-{z_0\over r+1} ,   \hspace{1 cm} & \textnormal{when} &   \hspace{1 cm}  \xi=-{1\over(r+1)}   \label{general self-similar 2}  
\end{eqnarray}
where $r=-1/(k+1)$. This expression was derived for an integer
$r$, however, one can verify  that expressions  (\ref{general self-similar}) and (\ref{general
self-similar 2}) are still solutions for a non-integer $r$ (different from 
$0$ and $-1$). 

\ack
The authors thank CNPq and Fapesp for support.

\end{document}